\documentclass[aps,prl,twocolumn,showpacs,reprint,superscriptaddress,longbibliography]{revtex4-1}
\usepackage{amsmath, amsfonts, amssymb, bm}
\usepackage{mathtools}
\usepackage[inline]{enumitem}
\usepackage{graphicx}
\usepackage{xcolor}
\usepackage[breaklinks,hypertexnames=false]{hyperref}
\hypersetup{colorlinks,linkcolor={magenta},citecolor={blue},urlcolor={blue}}

\usepackage[capitalize]{cleveref}
\usepackage{glossaries}
\glsdisablehyper     

\newacronym{mar}{MAR}{multiple Andreev reflections} \newacronym{sgs}{SGS}{subharmonic gap structure}
\newacronym{sss}{SSc}{spin-split (or Zeeman-split) superconductors}

\begin{document}

\title{Spin-polarized multiple Andreev reflections in spin-split superconductors}

\newcommand{\tianjin}{Center for Joint Quantum Studies and Department of Physics,
	Tianjin University, Tianjin 300072, China}
\newcommand{\aalto}{Department of Applied Physics,
	Aalto University, 00076 Aalto, Finland}
\newcommand{\nagoya}{Department of Applied Physics,
	Nagoya University, Nagoya 464-8603, Japan}

\author{Bo Lu}
\affiliation{\tianjin}

\author{Pablo Burset}
\affiliation{\aalto}

\author{Yukio Tanaka}
\affiliation{\nagoya}

\date{\today}

\begin{abstract}
We study the transport properties of a voltage biased contact between two spin-split superconductors separated by an insulating barrier of arbitrary transparency.
At low transparency, the contribution of multiple Andreev reflections leads to a subharmonic gap structure that crucially depends on the amplitude and relative angle of the spin-splitting fields of each superconductor.
For non-collinear fields, we find an interesting even-odd effect on the bound states within the gap, where the odd order multiple Andreev reflections split, but the even order ones remain at their expected positions.
By computing the current-voltage characteristics, we determine the transparency required for the emergence of a subharmonic gap structure and show that the splitting of the odd bound states is associated with different threshold energies of spin-polarized Andreev processes.
Our findings provide a tool to experimentally determine the amplitude and orientation of Zeeman fields in spin-split superconductors.
\end{abstract}

\maketitle

\begin{figure}
\includegraphics[width=0.95\columnwidth]{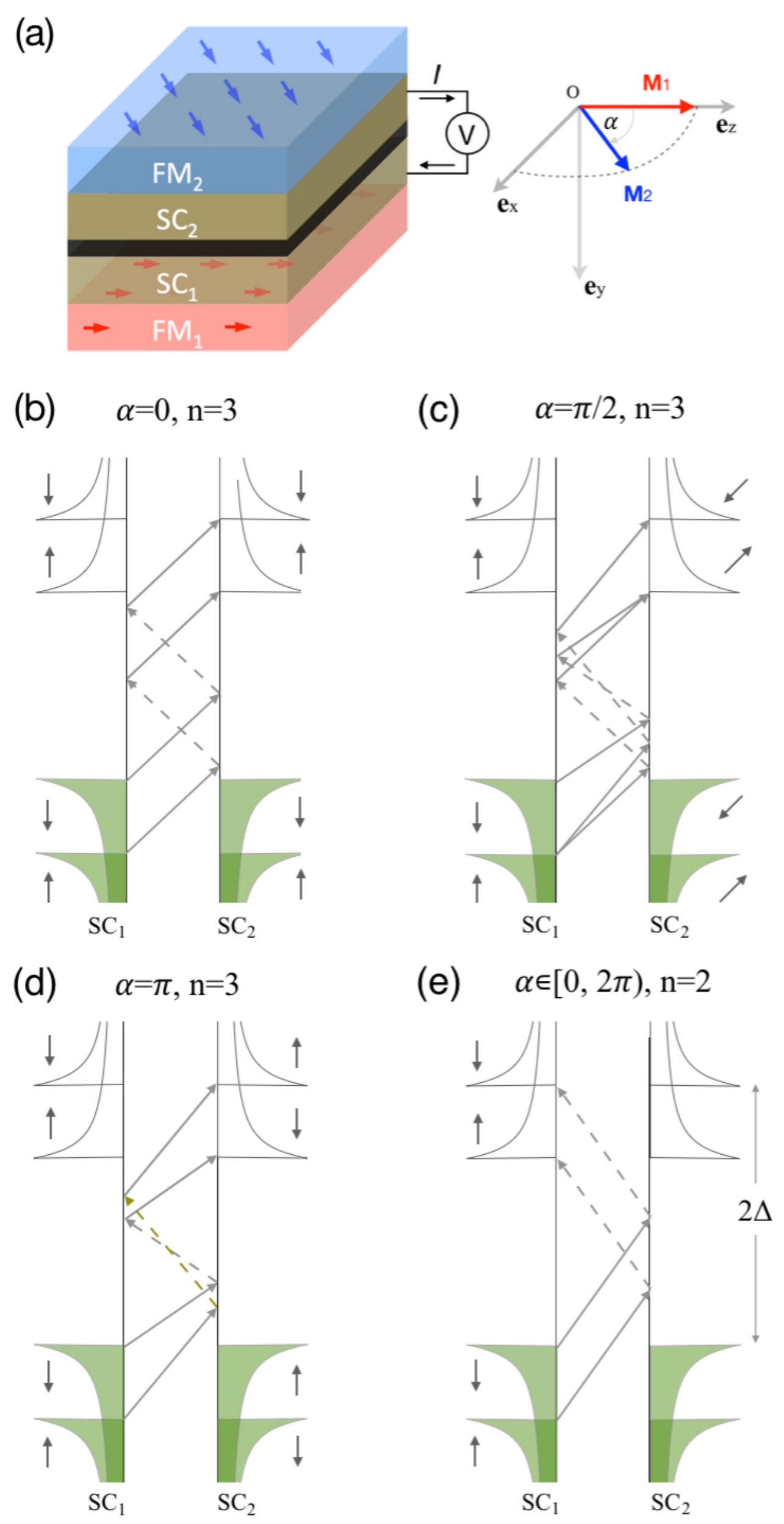}
\caption{(a) Schematic diagram of the multilayer FM-S-I-S-FM device used to develop a \gls{sss}-I-\gls{sss} Josephson junction, where I is an insulating barrier. The moments $\mathbf{M}_{L,R}$ of the FMs are confined to the $x-z$ plane, with a relative angle $\alpha$, while transport takes place along the $y$-direction.
(b)(c)(d) Schematic representation of the 3rd-order \gls{mar}. The superconducting gap is $\Delta$ and the energy bands are split by the Zeeman fields. Solid (dashed) line represent electron (hole) trajectories.
The relative angle $\alpha$ affects the positions of the threshold voltages for odd order \gls{mar}.
(e) 2nd-order \gls{mar}, with threshold voltage at $eV\!=\!\Delta$, is independent of $\alpha$.}
\label{fig:01}
\end{figure}

\emph{Introduction.---}
Hybrid structures between superconductors and magnetic materials reveal many interesting phenomena originated from the coexistence and interplay between ferromagnetism and superconductivity.
As a result, the new field of superconducting spintronics has emerged~\cite{Linder_2015,Eschrig_Review}, aiming at incorporating superconducting order into modern spintronic devices.
The creation of spin-triplet Cooper pairs and spin-polarized quasiparticles with long spin-coherence lengths~\cite{Yang_2010,Hubler_2012} suggest future applications based on a reliable and efficient manipulation of spin-polarized currents~\cite{Breunig_2018}.
In this context, superconductors with spin-split energy bands, commonly known as \gls{sss}, are attracting considerable interest~\cite{Bergeret_RMP2}. \gls{sss} can be realized either in a thin ferromagnet--superconductor (FM-S) junction via proximity effect~\cite{Kumar_1986,Bergeret_2001,Bergeret_RMP,Giazotto_2008}, cf. \cref{fig:01}(a), or in a thin-film superconductor subject
to a parallel (in-plane) magnetic field~\cite{Fulde_1970,Meservey_1971,Meservey_1973,Meservey_1973b}.
Highly spin-polarized currents can be generated in \gls{sss} hybrid junctions~\cite{Giazotto_2008,Linder_2008,Hubler_2012} and large thermoelectric effects have been predicted~\cite{Kuzmin_2012,Belzig_2013,Heikkilla_2014,Machon_2014,Bergeret_RMP2,Keidel_2019}.

The transport properties of hybrid junctions involving \gls{sss} has thus become a topic of fundamental interest in superconducting spintronics.
For example, the phase difference in Josephson junctions between \gls{sss} has been used as a source to manipulate the spin-polarized supercurrents~\cite{Tedrow_1987,Moodera_1991,Bergeret_2001,Xiong_2011,Bergeret_2012}. Recently, the tunneling quasiparticle current between two \gls{sss} linked by a spin-polarized barrier has been analyzed using quasi-classical Green function techniques, showing good agreement between theory and experiments~\cite{Rouco_2019}.
Here, we go one step further and analyze voltage biased \gls{sss} Josephson junctions of arbitrary transparency, where \gls{mar} play an important role, and find that the \gls{sgs} is very sensitive to the spin-splitting fields of the superconductors.

When two superconductors are in electric contact, \gls{mar}~\cite{Klapwijk_1982,Octavio_1983,Arnold_1987} take place at voltages $eV<2\Delta$, where $\Delta$ is the energy gap.
In a symmetric junction involving conventional superconductors, electrons or holes can undergo sequential Andreev reflections at the interface.
Due to the voltage bias, the quasiparticles will gain or lose an energy $eV$ as they travel across the interface, until escaping to the reservoirs for energies above the superconducting gap.
This phenomenon results in the so-called \gls{sgs}, a series of resonant conductance peaks at voltages $V_{n}\!=\!2\Delta /(en)$ in the current-voltage characteristics, where $n$ is an integer~\cite{Klapwijk_1982,Octavio_1983,Arnold_1987,Bratus_1995,Bardas_1995,Cuevas_1996}.
The peaks reveal the singular density of states at the superconducting energy gap edges, see \cref{fig:01}(b-e).
The study of the \gls{sgs} has proven to be useful for identifying properties of high-Tc superconductors~\cite{Poenicke_2002,Cuevas_2002}, topological superconductors~\cite{Badiane_2011,San_Jose_2013}, and other mesoscopic hybrid junctions~\cite{Cuevas_2001,Bobkova_2006,Cuevas_2006,Bobkova_2007,Dolcini_2008,Houzet_2009,Belzig_2014}.

In this paper, we study the transport properties in the \gls{mar} regime of a junction between two \gls{sss}, as depicted in \cref{fig:01}(a).
For simplicity, we only consider junctions where both superconductors have the same pair amplitude, $\Delta$, and strength of the spin-splitting field, $h$.
However, we allow the junction to be asymmetric by changing the relative orientation, $\alpha$, of the in-plane spin-splitting fields.
Without including phonon-induced spin-flip \cite{Fulde_1996}, we find that the \gls{sgs} due to \gls{mar} is highly tunable and presents a spin-dependent shift proportional to the Zeeman field $h$.
Additionally, depending on the relative orientation of the spin-splitting fields, we find an even-odd effect on the conductance peaks forming the \gls{sgs}.
\cref{fig:01} illustrates how \gls{mar} are modified by the spin-splitting fields. For $n$-order \gls{mar}, a quasiparticle undergoes $n-1$ Andreev reflections when transferring through the junction, and its spin is conserved since we are dealing with spin-singlet pairing states.
Importantly, while \gls{mar} processes with \textit{odd} order ($n$ an odd integer) involve quasiparticles transferred between superconductors, for \textit{even} order processes ($n$ an even integer), quasiparticles instead return to the same superconductor.
Therefore, only odd order \gls{mar} are sensitive to the relative orientation of the splitting fields.
The threshold energies for quasiparticles traveling from one side to the other are $2\Delta $ for parallel magnetization ($\alpha \!=\!0$) and $2\left( \Delta \!\pm\! h\right) $ for antiparallel magnetization ($\alpha \!=\!\pi $). Arbitrary values of $\alpha$ lead to spin-mixing and result in three possible channels $2\Delta $ and $2\left( \Delta \!\pm\! h\right) $.
Thus, the odd subharmonics become $2\Delta /n$ for $\alpha \!=\!0$ [see \cref{fig:01}(b)], $2\left( \Delta \!\pm\! h\right)/n$ for $\alpha \!=\!\pi $ [see \cref{fig:01}(d)], and $2\Delta /n$, $2\left( \Delta \!\pm\! h\right) /n$ for $\alpha \!\neq\! 0,\pi $ [see \cref{fig:01}(c)].
However, for even order processes, quasiparticles travel back to the same side, with spin conserved, and
thus the threshold energy gains or loses between the two gap edges are independent of the splitting field, as shown in \cref{fig:01}(e).
Therefore, the position of the conductance peaks for even subharmonics, $2\Delta/n$, is not altered by the spin-splitting fields.
The \gls{sgs} thus provides useful information about the strength and relative orientation of the spin-splitting fields at Josephson junctions between \gls{sss}.

\begin{figure*}
	\includegraphics[width=0.8\textwidth]{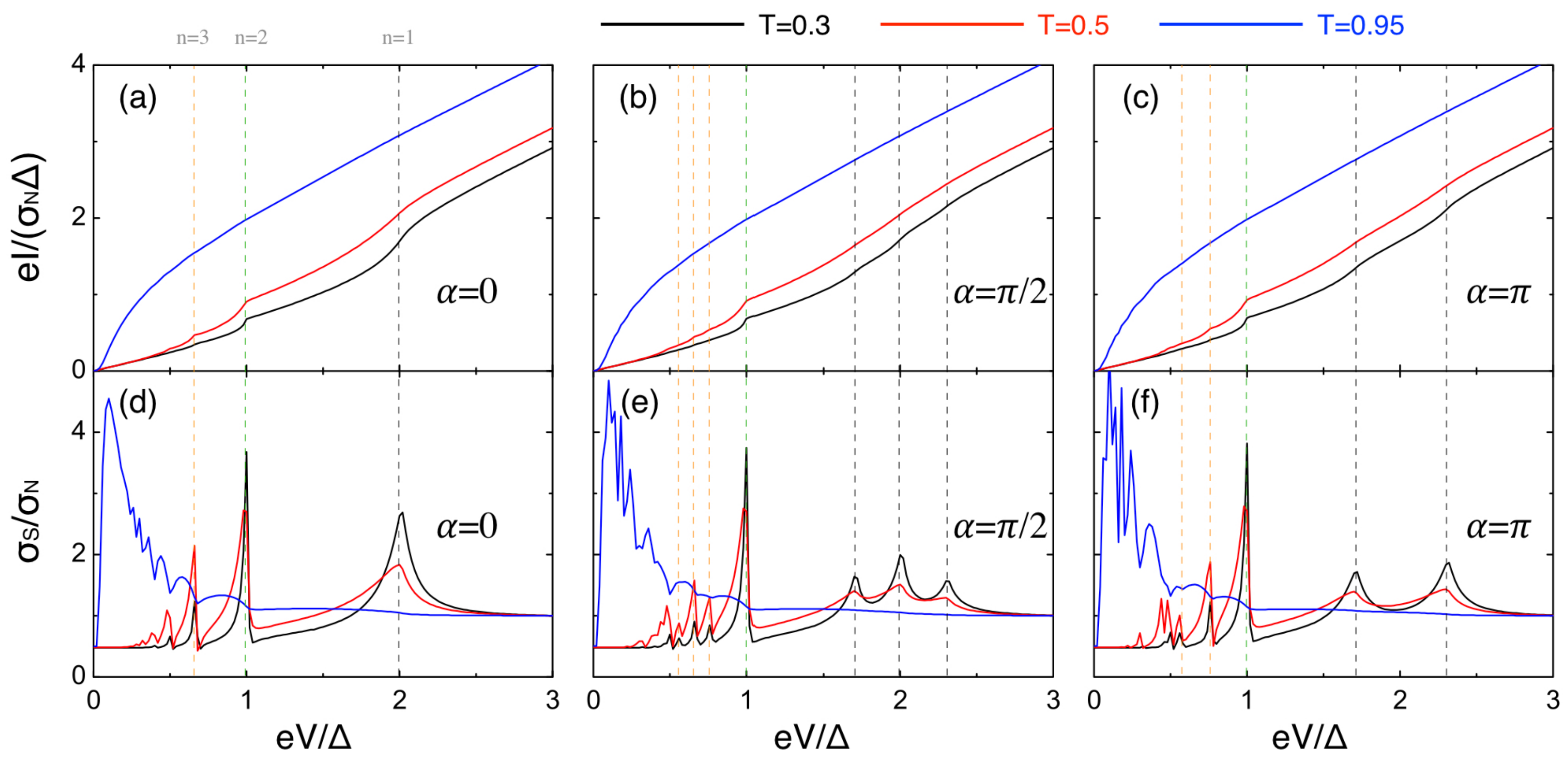}
	\caption{Electric current and differential conductance as a function of the voltage, for relative angles $\alpha\!=\!0,\pi/2,\pi$, and transmissions $T\!=\!0.95,0.5,0.3$. In all cases, $h/\Delta\!=\!0.15$. The dashed vertical lines show the $n$-th order \gls{sgs} at $n\!=\!1$ (black) $n\!=\!2$ (green) and $n\!=\!3$ (red). }
	\label{fig:02}
\end{figure*}

\emph{Model.---}
The system we study consists of two semi-infinite superconductors in a point contact geometry. Spin-splitting fields are induced on both sides via proximity effect to a FM region. The phenomenological Hamiltonian of the system is written as $\hat{H}\!=\!\hat{H}_{L}+\hat{H}_{R}+\hat{H}_{T}(t)$. $\hat{H}_{L}$ and $\hat{H}_{R}$ describe the bulk \gls{sss} on the left and right
side:
\begin{equation}
\hat{H}_{j=L,R}=\frac{1}{2}\sum_{k}\hat{\Psi}_{jk}^{\dagger }\left[
\begin{array}{cc}
H_{jk} & \hat{\Delta} \\
\hat{\Delta}^{\dag } & -H_{jk}^{T}%
\end{array}%
\right] \hat{\Psi}_{jk},  \label{eq1}
\end{equation}%
where $\hat{\Psi}_{jk}\!=\! [ \psi _{jk,\uparrow },\psi _{jk,\downarrow},\psi _{j(-k),\uparrow }^{\dagger },\psi _{j(-k),\downarrow }^{\dagger } ] ^{T}$ is the spinor in Nambu-spin space. The noninteracting Hamiltonian is
\begin{equation}
H_{jk}=\left[ k^{2}/(2m)-\mu \right] \hat{\sigma}_{0}-g\mu _{B}\mathbf{\hat{%
\sigma}\cdot M}_{j}.
\end{equation}
We assume that the orientation of the Zeeman fields lie in the $x-z$ plane and parametrize the moments as $\mathbf{M}_{L}\!=\!M\mathbf{e}_{z}$ and $\mathbf{M}_{R}\!=\!M\left[ \sin \alpha \mathbf{e}_{x}+\cos \alpha
\mathbf{e}_{z}\right] $, with $g$, $\mu _{B}$ and $\mathbf{M}_{L,R}$ the effective Land\'e $g$-factor, Bohr magneton, and induced magnetic fields, respectively.
Other spin orientations can be accounted for by an appropriate rotation without affecting our results.
The gap matrix in \cref{eq1} is $\hat{\Delta}\!=\!i\hat{\sigma}_{y}\Delta $ for spin-singlet $s$-wave pairing, where the Pauli matrices $\hat{\sigma}_{0,x,y,z}$ operate in spin space.
The tunneling term $\hat{H}_{T}(t)$ is given by
\begin{equation}
\hat{H}_{T}(t)=\frac{1}{2}\sum_{k,k^{\prime }}\,\left[ \hat{\Psi}%
_{Lk}^{\dagger }\,\hat{T}_{kk^{\prime }}(t)\,\hat{\Psi}_{Rk^{\prime }}+%
\mathrm{h.c.}\right] ,
\end{equation}%
with $\hat{T}_{kk^{\prime }}(t)\!=\!t_{kk^{\prime }}\hat{\tau}_{z}\,e^{i\chi (t) \hat{\tau}_{z}/2}$ and $\hat{\tau}_{x,y,z}$ the Pauli matrices in particle-hole space.
In the presence of a voltage bias $V$, the superconducting phase difference $\chi (t)\!=\!2\omega_{J}t$ is time-dependent, with $\omega_{J}=2eV$ the Josephson frequency. The resulting time-dependent current follows the Josephson relation $\Delta \rightarrow \Delta e^{i2eVt}$~\cite{Josephson_1965}, with $I\left( t\right) =\sum_{n}I_{n}e^{in\omega_{J}t}$.
For simplicity, we set $t_{kk^{\prime}}$ to be $t_{kk^{\prime }}\!=\!t_{0}\delta _{k,k^{\prime }}$ and denote $\hat{T}(t)\!=\!t_{0}\hat{\tau}_{z}\,e^{i\chi (t)\hat{\tau}_{z}/2}$.
Following the quasi-classical approximation~\cite{Lu_2016,Burset_2017}, we average the summation over channels $k$ without affecting the characteristics of the \gls{sgs}, but simplifying calculations.
By the so-called $\xi $-integration, the retarded/advanced Green function $\mathcal{\hat{G}}_{j}^{r/a}\left( \omega
\right) $ in the bulk state adopts the form \cite{LinderOdd_2015}
\begin{equation}
\mathcal{\hat{G}}_{L}^{r/a}=\left[
\begin{array}{cccc}
g_{\uparrow }^{r/a} & 0 & 0 & f_{\uparrow }^{r/a} \\
0 & g_{\downarrow }^{r/a} & f_{\downarrow }^{r/a} & 0 \\
0 & f_{\downarrow }^{r/a} & g_{\downarrow }^{r/a} & 0 \\
f_{\uparrow }^{r/a} & 0 & 0 & g_{\uparrow }^{r/a}%
\end{array}%
\right] \triangleq \left[
\begin{array}{cc}
\mathring{\Upsilon}_{11}^{r/a} & \mathring{\Upsilon}_{12}^{r/a} \\
\mathring{\Upsilon}_{21}^{r/a} & \mathring{\Upsilon}_{22}^{r/a}%
\end{array}%
\right] ,
\end{equation}
where
\begin{align}
\mathcal{\hat{G}}_{R}^{r/a}={}&\hat{U}\mathcal{\hat{G}}_{L}^{r/a}\hat{U}^{\dag
}\triangleq \left[
\begin{array}{cc}
\mathring{\Gamma}_{11}^{r/a} & \mathring{\Gamma}_{12}^{r/a} \\
\mathring{\Gamma}_{21}^{r/a} & \mathring{\Gamma}_{22}^{r/a}%
\end{array}%
\right] ,
\\
g_{\sigma }^{r/a}\left( \omega \right) & =\frac{-\left( \omega +\zeta
	_{\sigma }h\pm i0^{+}\right) }{W\sqrt{\Delta ^{2}-\left( \omega +\zeta
		_{\sigma }h\pm i0^{+}\right) }},
\\
f_{\sigma }^{r/a}\left( \omega \right) & =\frac{\Delta}{W\sqrt{\Delta
		^{2}-\left( \omega +\zeta _{\sigma }h\pm i0^{+}\right) }},
\end{align}
with $\zeta _{\sigma =\uparrow,\downarrow } \!=\!\pm 1$ and $\hat{U} \!=\! e^{-i\hat{\sigma}_{y}\alpha/2} \hat{\tau}_{0}$.
The magnitude of the proximity-induced spin-splitting fields is $h\!=\!g\mu _{B}M$ and $W$ is a band parameter \cite{Cuevas_2001}.
We choose the magnetic field below the so-called Chandrasekhar-Clogston limit~\cite{Chandrasekhar_1962,Clogston_1962}, $h_\text{max}\!=\!\Delta_{0}/\sqrt{2}$.
The Keldysh Green function $\mathcal{\hat{G}}_{j=L,R}^{K}\left( \omega \right) $ is given by $
\mathcal{\hat{G}}_{j}^{K}\left( \omega \right) \!=\! [ \mathcal{\hat{G}}_{j}^{r}\left( \omega \right) -\mathcal{\hat{G}}_{j}^{a}\left( \omega \right) ] \tanh [ \omega /\left( 2k_{B}T\right) ] $~
\footnote{The temperature dependence of the superconducting gap function is approximated as $\Delta \left( T\right) \!=\! \Delta _{0}\tanh \left( 1.74\sqrt{T_{c}/T-1}\right) $, with $\Delta _{0}\!=\!1.76k_{B}T_{c}$ and $T_{c}$ the critical temperature. We take a sufficient low temperature $k_{B}T\!=\!10^{-3}\Delta_{0} $ in our calculation and set the band parameter $W\!=\!1$ without qualitatively influencing our results.}.

\begin{figure*}
	\includegraphics[width=0.9\textwidth]{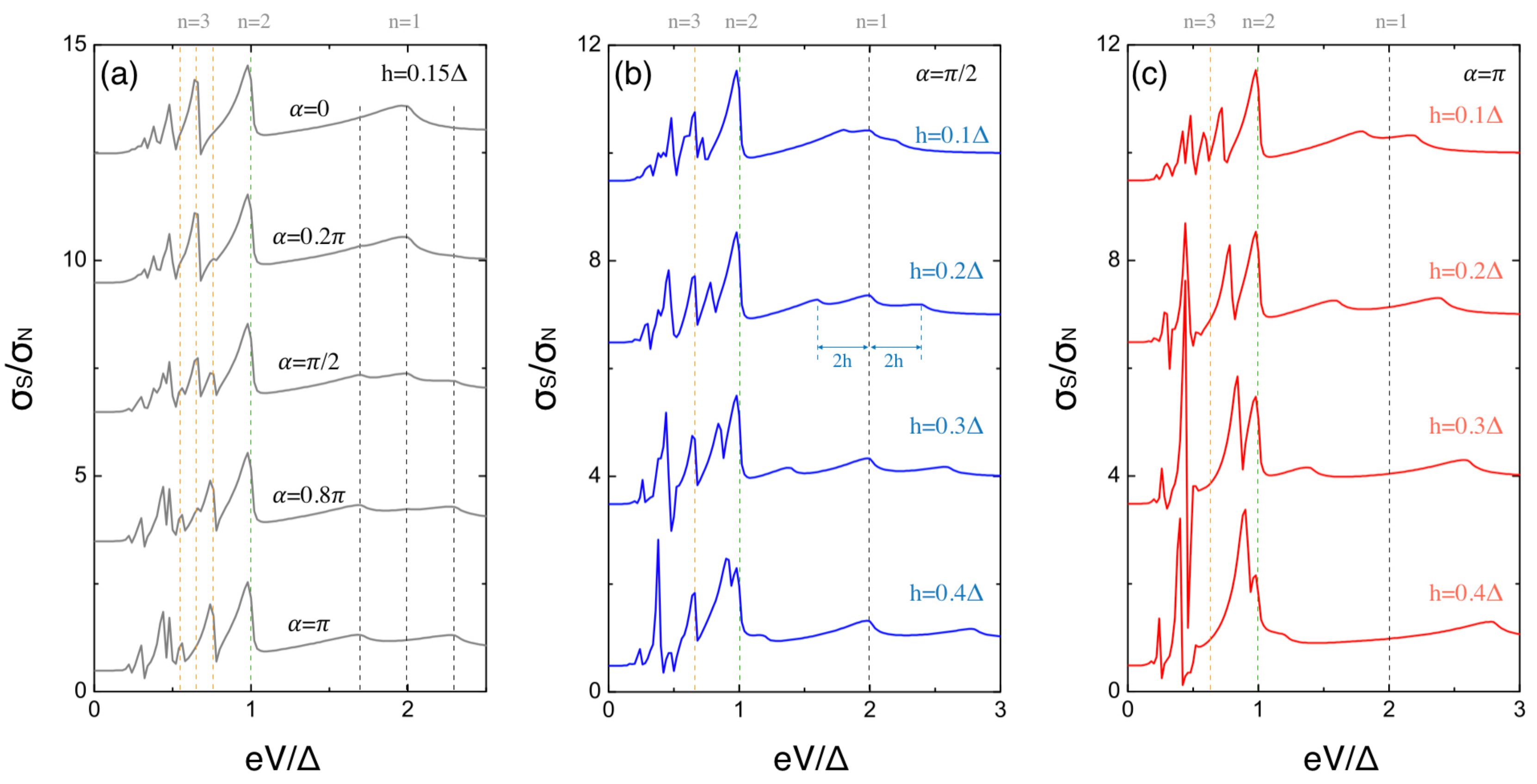}
	\caption{Normalized differential conductance as a function of the voltage.
		(a) Plot of the conductance for several relative angles, with fixed $h/\Delta\!=\!0.15$.
		(b,c) Plot of the conductance for various spin-splitting fields, with fixed $\alpha\!=\!\pi/2$ for (b) and $\alpha\!=\!\pi$ for (c).
		In all cases, the curves have been vertically displaced for clarity and the dashed vertical lines show the $n$-th order \gls{sgs} at $n\!=\!1$ (black) $n\!=\!2$ (green) and $n\!=\!3$ (red). }
	\label{fig:03}
\end{figure*}

Following the transfer-matrix approach and double Fourier
transformation~\cite{Cuevas_1996,Cuevas_2001,Dahm_2007}, we express the charge current in the following form:
\begin{equation}
\mathcal{I}\left( t\right) =e\nu \left( 0\right) v_{F}\sum\limits_{m}\int
\frac{d\omega }{8\pi }Tr\left[ \hat{\tau}_{z}\mathcal{I}_{m}\right]
e^{im\omega _{J}t/2},  \label{current}
\end{equation}
with $\nu \left( 0\right)$ the density of states at the Fermi level in the normal state. The components $\mathcal{I}_{m}$ are defined as
\begin{align}
\mathcal{I}_{m}={}&\sum_{n}\mathcal{\hat{G}}_{R,0}^{r}\left[ \mathcal{T}_{n0}^{a}\right] ^{\dag }\mathcal{\hat{G}}_{L,n}^{K}\mathcal{T}_{nm}^{a}
+
\mathcal{T}_{0n}^{r}\mathcal{\hat{G}}_{R,n}^{K}\left[ \mathcal{T}_{mn}^{r} \right] ^{\dag }\mathcal{\hat{G}}_{L,m}^{a}
\nonumber \\
& -
\left[ \mathcal{T}_{n0}^{a} \right] ^{\dag }\mathcal{\hat{G}}_{L,n}^{K}\mathcal{T}_{nm}^{a}\mathcal{\hat{G}}_{R,m}^{a}
-
\mathcal{\hat{G}}_{L,0}^{r}\mathcal{T}_{0n}^{r}\mathcal{\hat{G}}_{R,n}^{K}\left[ \mathcal{T}_{mn}^{r}\right] ^{\dag },
\end{align}
with $\mathcal{\hat{G}}_{j,n}^{r,a,K} \!=\!\mathcal{\hat{G}}_{j}^{r,a,K}\left( \omega +n\omega _{J}/2\right) $.
The dc component of the current corresponds to the $m\!=\!0$ harmonic in \cref{current}. Experimentally, it relates to the average electric current in the long time limit.
The transfer-matrix satisfies $\mathcal{T}_{nm}^{r/a}\left( \omega \right) \!=\! \mathcal{T}_{n-m,0}^{r/a}\left( \omega +m\omega _{J}/2\right) $ and can be determined by the recursive relation
\begin{align}\label{eq4}
\mathcal{\hat{T}}_{nm}^{r/a}= {}& t_{0}
\left[\begin{array}{cc}
\hat{\sigma}_{0} & \mathbf{0} \\
\mathbf{0} & \mathbf{0}%
\end{array} \right] \delta _{n,-1}
+
t_{0}\left[\begin{array}{cc}
\mathbf{0} & \mathbf{0} \\
\mathbf{0} & -\hat{\sigma}_{0}%
\end{array} \right] \delta_{n,1}
\\
& +
\hat{\epsilon}_{n}^{r/a}\mathcal{\hat{T}}_{nm}^{r/a}
+
\hat{V}_{n,n+2}^{r/a}\mathcal{\hat{T}}_{n+2,m}^{r/a}
+
\hat{\Lambda}_{n,n-2}^{r/a}\mathcal{\hat{T}}_{n-2,m}^{r/a} , \nonumber
\end{align}
with
\begin{align*}
\hat{\epsilon}_{n}^{r/a} ={}&t_{0}^{2}\left[
\begin{array}{cc}
\mathring{\Gamma}_{11,n+1}^{r/a}\mathring{\Upsilon}_{11,n}^{r/a} & \mathring{%
\Gamma}_{11,n+1}^{r/a}\mathring{\Upsilon}_{12,n}^{r/a} \\
\mathring{\Gamma}_{22,n-1}^{r/a}\mathring{\Upsilon}_{21,n}^{r/a} & \mathring{%
\Gamma}_{22,n-1}^{r/a}\mathring{\Upsilon}_{22,n}^{r/a}%
\end{array}%
\right] , \\
\hat{V}_{n,n+2}^{r/a} ={}&-t_{0}^{2}\left[
\begin{array}{cc}
\mathring{\Gamma}_{12,n+1}^{r/a}\mathring{\Upsilon}_{21,n+2}^{r/a} &
\mathring{\Gamma}_{12,n+1}^{r/a}\mathring{\Upsilon}_{22,n+2}^{r/a} \\
0 & 0%
\end{array}%
\right] , \\
\hat{\Lambda}_{n,n-2}^{r/a} ={}&-t_{0}^{2}\left[
\begin{array}{cc}
0 & 0 \\
\mathring{\Gamma}_{21,n-1}^{r/a}\mathring{\Upsilon}_{11,n-2}^{r/a} &
\mathring{\Gamma}_{21,n-1}^{r/a}\mathring{\Upsilon}_{12,n-2}^{r/a}%
\end{array}%
\right] .
\end{align*}

\cref{eq4} is numerically solved by introducing ladder operators $\mathcal{T}_{n,m}^{r/a} \!= \!z_{n,\pm }^{r/a}\mathcal{T}_{n\mp2,m}^{r/a}$ and using cut-off values $z_{n,\pm }^{r/a} \!= \!0$ for a sufficient large $\left\vert n\right\vert  \!= \!n_{N}$, where $\mathcal{T}_{\pm n_{N},0}^{r/a}$ are assumed to vanish. We normalize the current in units of $\sigma_{N}\Delta /e$, where $\sigma_{N}$ is the conductance when both electrodes are in non-superconducting states without magnetic elements.

\emph{Josephson dc current ---}
We now compute the current--voltage characteristics of a dc-biased contact between two \gls{sss}. As explained before, we set the gap of both superconductors to be the same, $\Delta$, and the spin-splitting field also equal, $h$.
We show in \cref{fig:02} the current and differential conductance, $\sigma_\text{S}$, for different values of the transmission, $T$, and relative orientation angle $\alpha$.
In the case of a high-transmission barrier, i.e., $T\!\lesssim\!1$, the current looks almost featureless, but its derivative displays small peaks due to \gls{mar}, which are more prominent at low voltages (blue lines in \cref{fig:02}). This is a result of the enhanced probability for Andreev reflections at high transmissions.
For lower transmissions, $T\!\lesssim\!0.5$, the subgap Andreev reflections are suppressed and thus the \gls{sgs} becomes more visible as kinks in the current and peaks in the differential conductance (red and black lines in \cref{fig:02}). These features are now more visible at higher voltages, due to the quasiparticle transfer at the gap edges in the tunneling regime.
These are common features of \gls{mar} in superconducting junctions; the novel effect comes from the angle $\alpha$ determining the relative orientations of the spin-splitting fields. \cref{fig:02} shows the three representative cases with $\alpha\!=\!0,\pi$ (parallel and anti-parallel) and $\alpha\!\neq\!0,\pi$ (non-collinear).
As observed in \cref{fig:02}(a,d), where the fields are parallel, the \gls{sgs} shows the usual distribution with peaks at $eV\!=\!2\Delta /n$. This result is similar to previous works~\cite{Cuevas_1996} without induced spin-split field in superconductors.
By contrast, when the fields are anti-parallel, the odd order \gls{mar} at $eV\!=\!2\Delta /n$ disappear while new peaks at $eV\!=\!2(\Delta \pm h)/n$ emerge ($n$ is an odd integer), see \cref{fig:02}(c,f).
Finally, when the magnetic field is non-collinear, the odd order \gls{mar} feature both types of peaks, at $eV\!=\!2\Delta /n$ and at $eV\!=\!2(\Delta \pm h)/n$, as shown in \cref{fig:02}(b,e).
Importantly, the splitting of the odd order peaks for $\alpha\!\neq\!0$ is directly proportional to the amplitude of the Zeeman field $h$.

From this result, we conclude that the odd \gls{sgs} are very sensitive to the configuration of the two spin-split fields. Thus, the $I-V$ characteristics could provide a way to measure the amplitude and relative orientation of the spin-splitting fields.
To clearly show these dependence on the spin-split fields, we plot the normalized differential conductance for a junction with average transmission $T\!=\!0.6$ in \cref{fig:03}.
First, we fix the magnitude of $h$ and change the relative angle in \cref{fig:03}(a). The odd order \gls{mar} split into three peaks when $\alpha\!\neq\!0$. The peaks at $eV\!=\!2(\Delta \pm h)/n$ reach their maximum as $\alpha $ approaches $\pi $. By contrast, the peaks at $eV\!=\!2\Delta /n$ are significantly reduced until they disappear for $\alpha\!=\!\pi $. Importantly, the even order \gls{mar} remain unchanged as $\alpha $ varies.
Next, we show in \cref{fig:03}(b)(c) the \gls{sgs} for several values of $h$ in the representative cases of non-collinear ($\alpha\!=\!\pi/2$) and anti-parallel ($\alpha\!=\!\pi$) fields.
The splitting of the odd order \gls{mar} becomes wider as $h$ increases. This result indicates the great tunability of the \gls{sgs} by both the relative angle and the magnitude of spin-split field. And the tunability is attributed to the spin-polarization of odd order \gls{mar}.

\emph{Conclusions ---}
We have theoretically analyzed the \gls{sgs} in a Josephson junction between two spin-split superconductors with arbitrary amplitude and orientation of their Zeeman fields. Our results show an interesting even-odd effect of the bound states within the gap. The \gls{sgs} induced by odd \gls{mar} are split by the Zeeman fields and they are strongly influenced by their relative angle. The \gls{sgs} from even \gls{mar}, instead, are independent of the spin-splitting fields.
The analysis of the \gls{sgs} is thus a useful tool to determine with great precision the magnitude and relative orientation of the Zeeman fields in experiments.
We remark that, although we consider a simplified single-channel superconducting weak link model, the splitting of odd-order \gls{mar} resonances is also present in common setups between spin-split superconductors, such as in a two-dimensional planar junction.

\emph{Acknowledgments ---}
We thank J. Cayao, S. Suzuki, S. Tamura and A. Yamakage for insightful discussions.
We acknowledge support from the Horizon 2020 research and innovation programme under the Marie Sk\l odowska-Curie Grant No. 743884 and the Academy of Finland (project 312299); Topological Material Science (Grants No. JP15H05851, No. JP15H05853, and No. JP15K21717); Grant-in-Aid for Scientific Research B (Grant No. JP18H01176) from the Ministry of Education, Culture, Sports, Science, and Technology, Japan (MEXT).

\bibliography{SpinSC}

\end{document}